\documentclass[12pt,a4paper]{article}
\usepackage{amsmath,amsfonts,graphics,epsfig,amssymb}
\begin{document}
\title{%
Muon content of ultra-high-energy air showers:
Yakutsk data versus simulations
}
\author{
A.V.~Glushkov, I.T.~Makarov, M.I.~Pravdin, I.E.~Sleptsov,
\\
(The Yakutsk EAS Array)
\\
\small \em
Yu.G.~Shafer Institute of Cosmophysical Research and
  Aeronomy,
  Yakutsk 677980, Russia\\
D.S.~Gorbunov,
G.I.~Rubtsov,
S.V.~Troitsky
\\
\small \em Institute for Nuclear Research of the Russian Academy of
Sciences,\\
\small \em
60th October Anniversary Prospect 7a, Moscow 117312 Russia
}
\date{}
\maketitle
\vspace{-12mm}
\begin{abstract}
We analyse a sample of 33 extensive air showers (EAS) with estimated
primary energies above $2\cdot 10^{19}$~eV and high-quality muon data
recorded by the Yakutsk EAS array.  We compare, event-by-event, the
observed muon density to that expected from CORSIKA simulations for
primary protons and iron, using SIBYLL and EPOS hadronic interaction
models. The study suggests the presence of two distinct hadronic
components, ``light'' and ``heavy''.
Simulations with EPOS are in a good
agreement with the expected composition in which the light component
corresponds to protons and the heavy component to iron-like nuclei.
With SYBILL, simulated muon densities for iron primaries are a factor of
$\sim 1.5$ less than those observed for the heavy component, for the same
electromagnetic signal.
Assuming two-component proton-iron composition and the EPOS model, the
fraction of protons with energies
$E>10^{19}$~eV is $0.52 ^{+0.19}_{-0.20}$ at 95\% confidence level.
\end{abstract}

Number of muons in extensive air showers (EAS) is used as an estimator
of the primary composition of ultra-high-energy cosmic rays detected
by surface arrays. Precision of muon-based composition
studies is limited by their sensitivity to hadronic interaction models
which incorporate extrapolation of experimental data to kinematic
regions never tested in a laboratory experiment. On the other hand,
the responses of the ground-based cosmic-ray detectors to different
EAS components are different, and variations in the muon content may
affect the relation between the signal recorded on the ground and the
inferred energy of the primary particle. This is important in
particular for detectors sensitive to the muon component
(e.g.\ the surface detector of the Pierre Auger Observatory).

In this {\it Letter} we compare the observed and simulated muon contents of EAS
detected by the Yakutsk array~\cite{YakutskExperiment}. This experiment is
currently the only one capable of detecting air showers initiated by
particles of $E \gtrsim 10^{19}$~eV and equipped with muon detectors.
Large area and high saturation threshold of each detector station make it
possible to obtain high-quality muon data~\cite{Yakutsk_mu}.

Recently, the Pierre Auger Observatory (PAO) collaboration
reported~\cite{Engel} an excess of muons as compared to simulations with
the QGSJET~II hadronic interaction model~\cite{QGSJET}. Since PAO is not
equipped with muon detectors, indirect methods were used.  The same result
has been previously reported~\cite{Yakutsk-muon-excess} by the Yakutsk
collaboration (by making use of older hadronic models). Here, we perform
a direct and detailed study of this effect for each individual shower in a
high-quality sample and for the sample as a whole, in the frameworks of
two different hadronic interaction models, SIBYLL~\cite{SIBYLL} and
EPOS~\cite{EPOS} (the muon content of showers simulated with QGSJET~II is
between these two~\cite{Pierog:2007zz}). We use a precise statistical method~\cite{composition}
to construct the distributions of muon densities simulated for individual
events, which are then compared to the observed data (see
Refs.~\cite{GammaLimit,Yak-gamma} for other applications of the method).

In our study, we use a sample of 33 events with reconstructed
energies above $2\cdot 10^{19}$~eV, zenith angles up to 45$^\circ$,
core location inside the array and high-quality muon data
recorded. The latter criterion means that we require at least three
operating muon detectors at distances between 400~m and 2000~m from the
shower axis, which allows one to reconstruct the lateral distribution
of the muon density. For each of the events, we simulated a library of
showers with different primary energies but with the same arrival
direction as observed. The simulations were performed with CORSIKA
6.611~\cite{corsika} using FLUKA 2006.3~\cite{fluka} as a low-energy
hadronic interaction model and either SIBYLL 2.1~\cite{SIBYLL} or EPOS
1.61~\cite{EPOS} as a high-energy model. For SIBYLL, multisampling
$(20\times 10^{-4})$ was used to suppress artificial fluctuations due
to thinning~\cite{our-thinning}. For the case of EPOS, we used
thinning ($10^{-5}$) with weights limitations~\cite{Thin} to save
computational time; more artificial fluctuations are expected in this
case. The responce of the scintillators was simulated with GEANT in
Ref.~\cite{YakutskGEANT}.

The Yakutsk collaboration uses the signal density at 600~m from the shower
core, $S(600)$, to estimate the energy $E_0$ of the primary
particle~\cite{Yakutsk_Eest}. The relation between $S(600)$ and $E_0$ was
obtained by making use of the constant intensity cuts method (to relate
$S(600)$ in inclined and vertical showers) and of the calibration by
Cherenkov light (to relate $S(600)$ and $E_0$ for vertical showers). In
this study, however, we do not use the estimated energy\footnote{Except
for the determination of the data set.} and hence do not use these relations;
instead, we use $S(600)$ measured in individual showers to compare real
and simulated events. To obtain the value of $S(600)$ in simulated
showers, we fit the lateral distribution function with the same
formula~\cite{Yakutsk_Eest} as used in processing of real data. We use the
muon density at 1000~m from the shower axis, $\rho _\mu (1000)$, as the
estimator of the muon content.

Making use of the method described in detail in
Ref.~\cite{composition}, we selected simulated showers whose $S(600)$
are  consistent with the observed one
taking into account experimental errors
in the determination of this parameter, and obtained the probability
distribution $f(\rho _\mu ^{\rm sim})$ to have $\rho _\mu (1000) =
\rho _\mu ^{\rm sim}$ in a simulated shower arriving from the same
direction and having the same $S(600)$ as the real one.  This
procedure was repeated for each event for assumed proton and iron
primaries. We used from 400 to 1000 simulated showers for each observed
event depending on its zenith angle. These distributions were analyzed
to estimate the ratios $\eta_{\rm (p,Fe)}\equiv\rho _\mu ^{\rm obs}/\rho
_\mu ^{\rm sim~(p,Fe)}$ for each shower. Averaged over all showers they
read
\begin{align}
{\rm SIBYLL:}&
~~~\eta_{\rm p}=1.92^{+0.24}_{-0.10} ,~~~
\eta_{\rm Fe}=1.00^{+0.09}_{-0.04},
\label{2*}\\
{\rm EPOS:}&
~~~\eta_{\rm p}=1.14^{+0.16}_{-0.06},~~~
\eta_{\rm Fe}=0.66^{+0.06}_{-0.03}.
\label{2**}
\end{align}
(only statistical errors are shown). 
The {\em average} composition is therefore
iron-like assuming the SIBYLL interaction
model and slightly heavier than protons assuming EPOS.

\begin{figure}
\includegraphics[width=0.5\columnwidth]{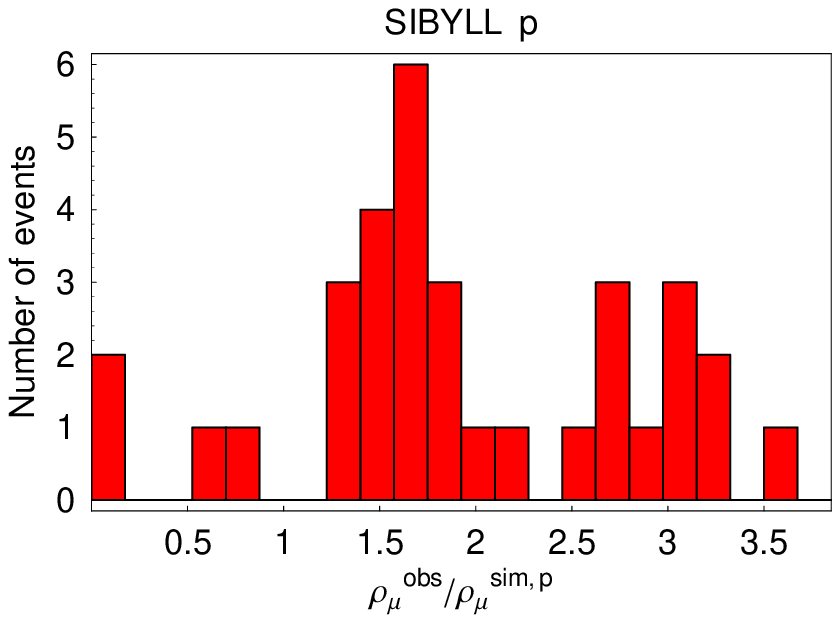}
\includegraphics[width=0.5\columnwidth]{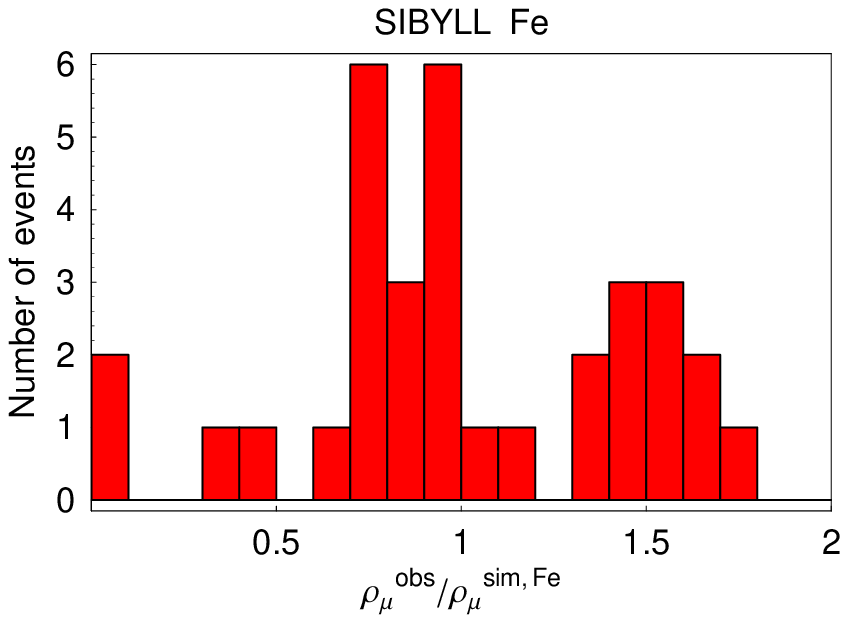}
\caption{
\label{fig:etaSIBYLL}
Distributions of $\eta _{\rm p}$ and $\eta _{\rm Fe}$ with SIBYLL.
}
\end{figure}

\begin{figure}
\includegraphics[width=0.5\columnwidth]{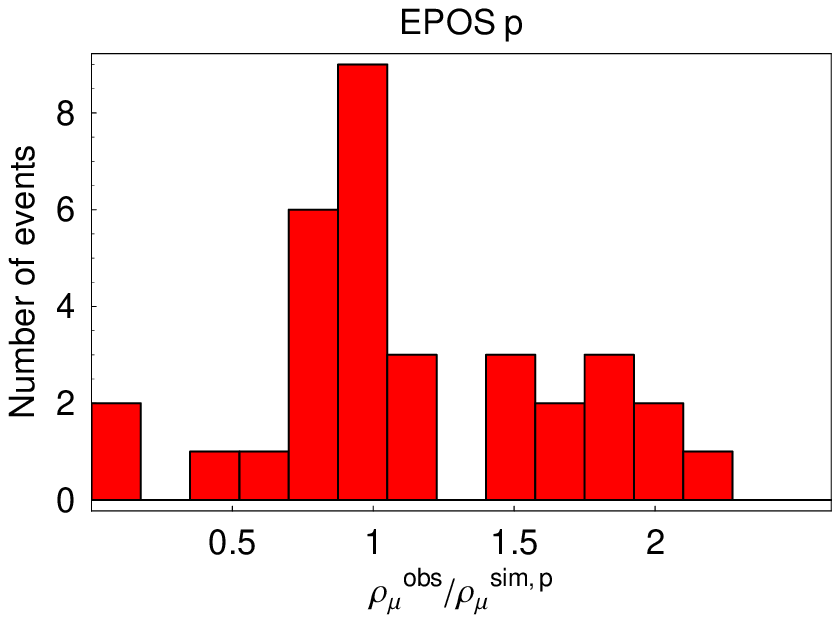}
\includegraphics[width=0.5\columnwidth]{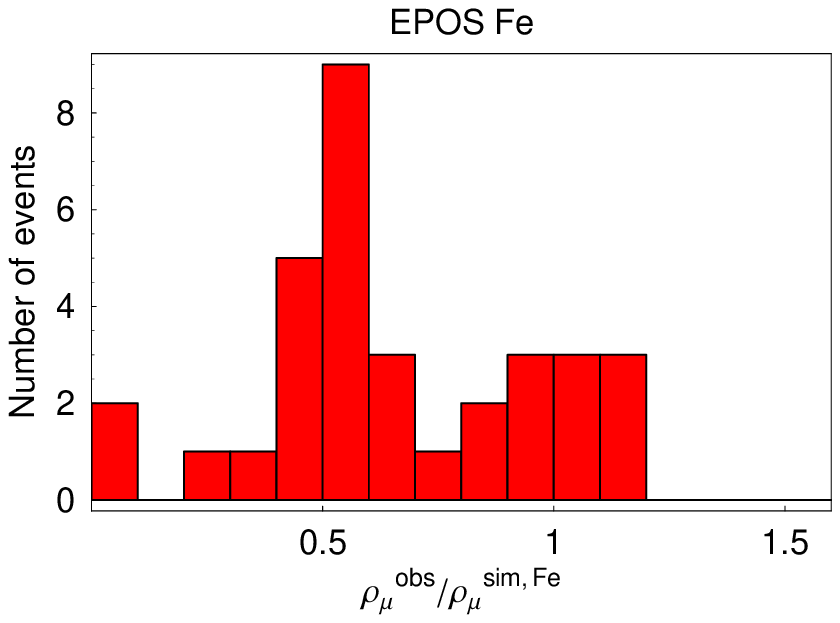}
\caption{
\label{fig:etaEPOS}
Distributions of $\eta _{\rm p}$ and $\eta _{\rm Fe}$ with EPOS.
}
\end{figure}

However, a glance at the distribution of $\eta $
calculated event-by-event (Figs.~\ref{fig:etaSIBYLL}, \ref{fig:etaEPOS})
suggests that averaging is not a proper way to interpret the data which
are actually trimodal.

The two (almost) muonless events may be interepreted as
photons~\cite{Yak-gamma}; here we concentrate on the rest of the
sample.  One is tempted to interpret two other peaks as a signature of
the mixed composition with the low-muon-density peak corresponding to
protons and the second peak corresponding to heavy nuclei. As we will
discuss elsewhere, this picture is indeed expected for UHECRs of
astrophysical origin because the mean free path of intermediate-mass
nuclei in the bath of cosmic background radiation is considerably
shorter, see for instance Ref.~\cite{Allard:2005ha}.

The intuitive requirement that the cosmic-ray hadrons are stable nuclei
with atomic numbers $1\le Z\le 56$ appears inconsistent with the results of
the SYBILL simulations (see Fig.~\ref{fig:etaSIBYLL}). Following
Ref.~\cite{Engel}, one has to assume a multiplicative correction factor for
the simulated muon density which should be about 1.5 to satisfy this
requirement. On the other hand, results of the EPOS simulations agree with
expectations within statistical uncertainties. Our results are based on
the simulation of both the signal density $S(600)$ and the muon density
$\rho _\mu (1000)$ and are thus sensitive not only to the hadronic model
but also to the model of electromagnetic interactions
(EGS4~\cite{Nelson:1985ec}). In principle, systematic underestimation of
the electromagnetic signal by EGS4 may explain the SYBILL result (then
EPOS simulates too high muon densities). Here,
we accept the option that both EGS4 and EPOS are correct
and do not consider the sample of showers simulated with SIBYLL in the
rest of the study.

\begin{figure}
\includegraphics[width=\columnwidth]{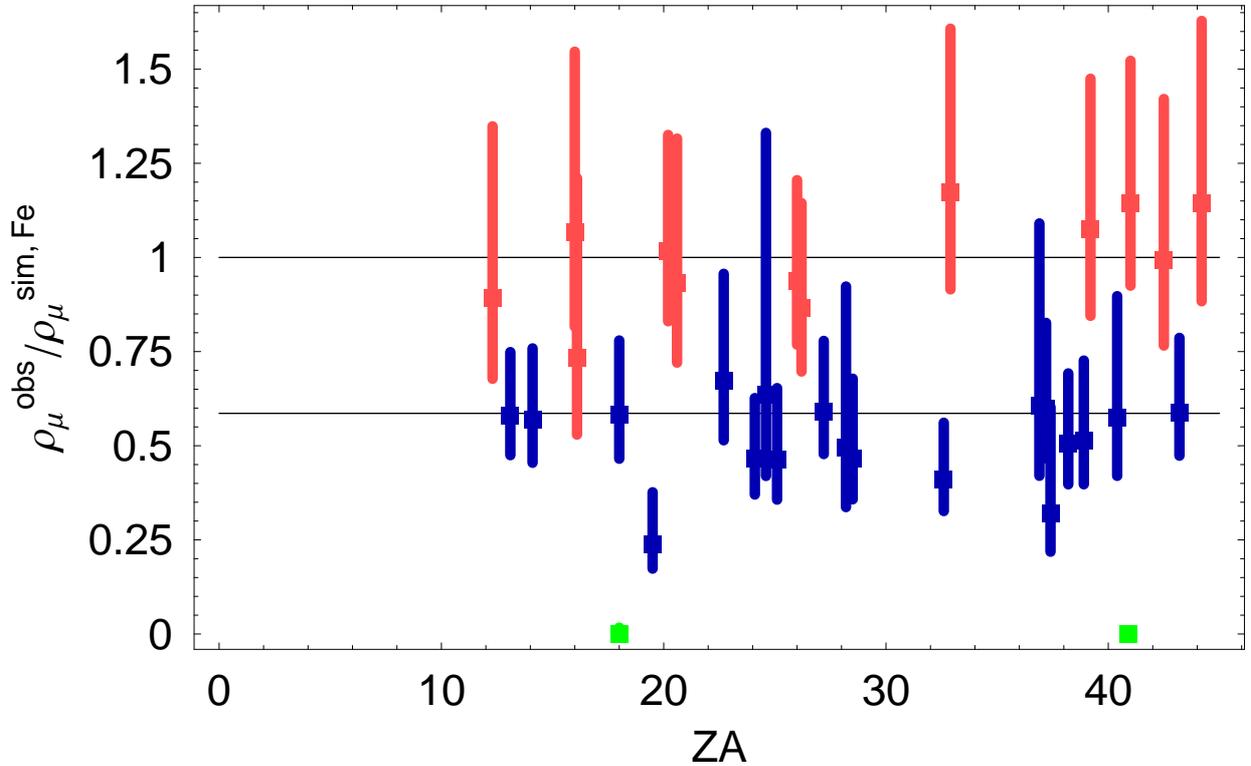}
\caption{
\label{fig:3modal}
The ratios $\eta^{\rm Fe}=\rho _\mu ^{\rm obs}/\rho _\mu ^{\rm sim, Fe} $
of observed and simulated muon densities at 1000~m from the shower core
versus the zenith angle. Central values and 68\%CL error bars (both
statistical and systematic errors are included) were obtained for
each event from the actual distributions of iron-induced showers simulated
with the EPOS hadronic model for the arrival direction and $S(600)$
consistent with the observed ones. The error bars include the precision of
reconstruction of $S(600)$ and $\rho _\mu (1000)$ as well as intrinsic
shower-to-shower fluctuations.
Two muonless events ($\eta^{\rm Fe}=0$) are shown in light green; the rest
of events are separated into two groups, proton-like (dark blue) and
iron-like (light pink), based on event-by-event probabilities calculated
as in Ref.~\cite{composition}. The horisontal lines represent expected
average values of $\eta^{\rm Fe}$ for iron (1.00) and protons (0.59).
}
\end{figure}

The distribution of muon densities of simulated showers presented in
Fig.~\ref{fig:etaEPOS} clearly indicates the presence of heavy nuclei
in UHECRs. The limited statistics of our sample
gives no chance to distinguish imprints of
intermediate mass nuclei, e.g. CNO, and we leave this problem for future.
The present sample is fully consistent with
purely three-component (photons, protons and iron) primary
composition.
This fact is illustrated in Fig.~\ref{fig:3modal},
which demonstrates that most of our 33 events can be consistently
identified with photon, proton or iron primary.

One may estimate quantitatively the significance of bimodality seen in the
distribution of $\eta$,
Figs.~\ref{fig:etaSIBYLL}, \ref{fig:etaEPOS},
for all events except muonless ones. The simplest
measure of bimodality is the kurtosis $c_4$, the 4th moment of a
distribution. A slightly better measure suitable for non-symmetric
distributions is the so-called bimodality coefficient
$b=(1+c_3^2)/(c_4+3)$, where $c_3$ is the skewness (the 3rd moment) of a
distribution and we use the definition of $c_4$ which provides $c_4=0$
for the normal distribution. We simulated 10000 sets, each of 31 numbers
drawn from a normal distribution with the same mean and standard deviation
as in the real data for $\eta=\rho_\mu^{\rm (obs)}/\rho_\mu^{\rm (sim)}$
and estimated the fraction of simulated sets which have the same or
stronger bimodality than the data, using both $c_4$ and $b$. This fraction
amounts to a few per cent (the precise number depends on the interaction
model and assumed primary used to calculate $\rho_\mu^{(sim)}$, as well as
on the estimator, $c_4$ or $b$; the typical numbers are 0.03 to 0.05). We
therefore conclude that the bimodality is confirmed at the 95\% confidence
level.

As discussed in detail in Ref.~\cite{composition}, the method we use is
not sensitive to the relation between $S(600)$ and energy used for the
experimental energy determination: simulated showers are selected by the
value of $S(600)$. Thus a sample of showers with reconstructed energies
$E>2\cdot 10^{19}$~eV may be used to constrain the primary composition for
a different energy range, related to $S(600)$
by simulations. Motivated by the fact that a given $S(600)$ corresponds to
a lower energy of a primary of a shower simulated with EPOS as compared to
the Yakutsk energy reconstruction procedure (more details will be
discussed elsewhere), we use our sample to infer primary composition at
$E>10^{19}$~eV. The correction for the possible incompleteness of the
sample determined by a cut in the reconstructed energy (``lost
particles'', see Ref.~\cite{composition} for details) is taken into
account in the results presented below. Systematic uncertainties of the
approach have been discussed in
Refs.~\cite{composition,GammaLimit,Yak-gamma}.

Following the procedure of Ref.~\cite{composition},
we estimate the proton and iron fractions in the integral flux of cosmic
rays at $E>10^{19}$~eV assuming two-component proton-iron composition and
the EPOS interaction model.
The most probable proton fraction is $\epsilon_p\approx 0.52$ with the
allowed interval
\begin{equation*}
\label{new*}
0.32\le \epsilon _{\rm p} \le 0.71 ~~~( 95\% {\rm CL}),~~~~~
E>10^{19}~{\rm eV}\;,
\end{equation*}
which corresponds to the fraction of the heavy (iron-like) component
\begin{equation*}
\label{new**}
0.29\le \epsilon _{\rm Fe} \le 0.68 ~~~( 95\% {\rm CL}),~~~~~
E>10^{19}~{\rm eV}\;.
\end{equation*}
These results are in a good agreement with the
results~\cite{AugerXmax} of hybrid observations at PAO and in a worse
agreement with the results~\cite{HiResXmax} of HiRes.

One may in principle try to distinguish between various possible reasons
of the discrepancies between simulated (with SYBILL in our case) and
observed muon densities. In particular,
incorrect extrapolation of the normalization of the cross section and/or
multiplicity of hadronic interactions in a simulation code would probably
result in a discrepancy in the total muon number of a shower while
problems with the distribution of transverse momenta may show themselves
in deviation of the observed muon lateral distribution function (LDF) from
the simulated one.
A more detailed study is in
progress which will include the analysis of signals at each
particular detector without fitting LDF by any predetermined function, the
study of the QGSJET hadronic model and the analysis of the depth of
the maximal shower development expected from the present results.
The sample of events will be supplemented by inclined showers.

As may be inferred from Fig.~1 of Ref.~\cite{Engel} and from the
discussion there, the energies of iron-like nuclei would on average be
estimated by the PAO surface detector by some 15\% higher than the
energies of protons, for the same true energy of the primary. Given the
steeply falling UHECR spectrum and assuming the mixed composition favoured
by the present study, the samples of events with reconstructed energies
higher than a fixed one would be biased towards heavy primaries. This may
explain the results of Ref.~\cite{Engel} pointing to a somewhat stronger
underestimation of average muon content by simulations than indicated by
the results, Eqs.~(\ref{2*}), (\ref{2**}), of the present work.

To summarize, we compared the observed muon densities of UHE air showers
with those simulated with different hadronic models and for different
primaries. The mixed primary composition was resolved,
on an event-by-event basis, into two distinct, light (proton-like) and
heavy (iron-like) components. The results reported here, namely the
presence of the heavy and light components in the UHECR flux, consistency
of EPOS and failure of SIBYLL to reproduce correctly the muon content of
air showers, may have substantial consequences for understanding the origin
of UHECR and the physics of high-energy hadron collisions, as well as for
the energy estimation by surface detectors sensitive to muons (currently
operating Pierre Auger Observatory as well as SUGAR and Haverah Park
arrays).

We are indebted to Yu.~Bykov, V.~Rubakov and I.~Tkachev for helpful
discussions. This work was supported in part by the Russian Foundation
of Basic Research grants 07-02-00820 (INR team), 05-02-17857a (the
Yakutsk team), by the grants of the President of the Russian
Federation NS-7293.2006.2 (government contract 02.445.11.7370; INR
team), MK-2974.2006.2 (DG), NS-7514.2006.2 (government contract
02.120.11.7514; the Yakutsk team), by the Dynasty Foundation (GR) and
by the Russian Science Support Foundation (ST). Numerical part of the
work was performed at the computer cluster of the Theoretical Division
of INR RAS.


\end{document}